\def\year{2020}\relax
\documentclass[letterpaper]{article} 
\usepackage{aaai20}  
\usepackage{times}  
\usepackage{helvet} 
\usepackage{courier}  
\usepackage[hyphens]{url}  
\usepackage{graphicx} 
\usepackage{graphicx}  
\newtheorem{definition}{Definition}
\usepackage{amsmath,amssymb,amsfonts}
\usepackage{algorithmic}
\usepackage[nomain]{glossaries}
\usepackage{tkz-graph}
\usepackage[ruled ]{algorithm2e}

\newglossary{symbols}{sym}{sbl}{List of Abbreviations and Symbols}

\newglossaryentry{T}{type=symbols,name={\ensuremath{T}},sort=T,
	description={Temporal Graph}}
\newglossaryentry{Gi}{type=symbols,name={\ensuremath{G_i}},sort=gt,
	description={Temporal Graph Snapshot}}
\newglossaryentry{ImpAll}{type=symbols,name={$\mathcal{I}$},sort=impall,
	description={Set of Importance value for each window}}
\newglossaryentry{ImpI}{type=symbols,name={$I_i$},sort=impl,
	description={Importance of a window}}
\newglossaryentry{Fm}{type=symbols,name={$\digamma_{k}  $},sort=Fm,
	description={Expected Frequency of motif m in the temporal graph}}
\newglossaryentry{F}{type=symbols,name={$\digamma $},sort=F,
	description={Temporal Motif distribution for the given temporal graph}}

\tikzset{global scale/.style={
		scale=#1,
		every node/.style={scale=#1}
	}
}

\SetCommentSty{mycommfont}
\title{ITeM: Independent Temporal Motifs to Summarize and Compare Temporal Networks }
\author{Sumit Purohit, George Chin\\
	Pacific Northwest National Laboratory\\
	Richland, WA, USA\\
	\{Sumit.Purohit,George.Chin\}@pnnl.gov
	\And
	Lawrence B. Holder\\
	Washington State University\\
	Pullman, WA, USA\\
	holder@wsu.edu}
\begin{document}
\maketitle

\begin{abstract}
Networks are a fundamental and flexible way of representing various complex systems. Many domains such as communication, citation, procurement, biology, social media, and transportation can be modeled as a set of entities and their relationships. Temporal networks are a specialization of general networks where the temporal evolution of the system is as important to understand as the structure of the entities and relationships. 
We present the \textit{Independent Temporal Motif} (ITeM) to characterize temporal graphs from different domains. ITeMs can be used to model the structure and the evolution of the graph. For a given temporal graph, we produce a feature vector of ITeM frequencies and apply this distribution to the task of measuring the similarity of temporal graphs. We show that ITeM has higher accuracy than other motif frequency-based approaches. We define various metrics based on ITeM that reveal salient properties of a temporal network. We also present \textit{importance sampling} as a method for efficiently estimating the ITeM counts. We evaluate our approach on both synthetic and real temporal networks.
\end{abstract}

\section{Introduction}
Networks have been widely used to represent entities, relationships, and behaviors in many real-world domains including power grids \cite{chu2017complex}, social networks \cite{kumar2010structure}, microbial interaction networks \cite{shen2018high}, corporate networks \cite{takes2018multiplex}, the food web \cite{klaise2017origin}, and modeling adversarial activities \cite{cottam2018multi}. These complex systems do not show a temporal or structural continuum, but rather show a characteristic non-linear dynamic behavior \cite{ben2004complex,toroczkaicomplex}. Many salient properties of these systems can be described by different network metrics, measured on a global scale. These properties are not possible to measure in real time for many domains generating a constant stream of heterogeneous network channels. Count-based metrics such as the number of entities, the number of interactions, and the average connectivity of the entities in the network are important measures that represent the population and the interaction density of the entities involved in the network. However, these measures are limited in their ability to describe non-linear, localized, and dynamic properties of the systems. In order to uncover structural, temporal, and functional insights of complex systems, \textit{network motifs} have been used extensively in recent years as they provide a tractable approximation of the networks that can be measured and updated within given memory and compute constraints. Network motifs are patterns of interactions occurring in the complex system at a rate higher than those in a randomized network \cite{milo2002network}.

Extensive research has been done on the appropriate definition of \textit{network motifs} \cite{milo2002network,vazquez2004topological} and their application to various network analytical tasks such as: defining \textit{network backbone} \cite{cao2019motif}, clustering microbial interactions \cite{shen2018high}, and identifying the exchange of emotions in online communication networks \cite{kuvsen2019analysis}. Jin et al. (\citeyear{jin2007trend}) define \textit{TrendMotif} that describes a recurring subgraph of weighted vertices and edges in a dynamic network over a user defined period. The \textit{TrendMotif} can indicate the increasing and/or decreasing intervals for the weighted vertices or edges over the time period. Borgwardt et al. (\citeyear{borgwardt2006pattern}) extend pattern mining on static graphs to time series of graphs where each graph has the same set of vertices and observed addition and deletion of edges.


A \textit{temporal network} is a generalization of a static network that changes with time. Many system modeling approaches model time as an attribute of the entity or the interaction, which makes temporal graphs a special case of \textit{attributed graphs}. We interchangeably use \textit{network} and \textit{graph} in this paper. Incorporating time into static graphs has given rise to a new set of important and challenging problems that cannot be modeled as a static graph problem \cite{kovanen2011temporal,michail2016introduction}. Network motifs are also used to visualize and summarize large dynamic graphs \cite{liu2018graph}. TimeCrunch \cite{shah2015timecrunch} \cite{shah2017summarizing} discovers five different temporal patterns of some common substructures and summarizes the network in terms of a sequence of substructures that minimizes the Minimum Description Length (MDL) cost of describing the graph. Adhikari et al. (\citeyear{adhikari2017propagation}) use local substructures to condense a temporal network. Liu et al. (\citeyear{liu2013detecting}) propose a Bayesian framework to estimate the number of temporal motifs in communication networks. A majority of the prior research does not account for the temporal evolution of the motif. Recent work \cite{paranjape2017motifs} defines $\delta-$temporal motif as an elementary unit of the temporal network and provides a general methodology for counting such motifs. It computes the frequency of \textit{overlapping} temporal motifs, where one interaction can be part of more than one temporal motif. In a $\delta-$temporal motif, all the edges in a given motif have to occur inside the time period of  $\delta$ time units. Li et al. (\citeyear{li2018temporal}) propose temporal Heterogeneous Information Networks (HIN) and develop a set of algorithms to count HINs. Apar{\'\i}cio et al. (\citeyear{aparicio2018graphlet}) use \textit{orbit} transitions to compare a set of temporal networks. Dynamic Graphlet (DG) \cite{hulovatyy2015exploring} extends static graphlets to analyze structure and function of molecular network. DG distinguishes \textit{graphlet} from motif as induced subgraphs that are not defined based on the statistical significance of the substructure. DG defines orbit in a graphlet to measure automorphism in the graphlet. Sarkar et al. (\citeyear{sarkar2019understanding}) use the temporal motif to understand information flow in social networks.

We propose the \textbf{I}ndependent \textbf{Te}mporal \textbf{M}otif (ITeM) as the elementary building block of temporal networks. In contrast to the related work, ITeMs are edge-disjoint temporal motifs that provide insight about the temporal evolution of a graph, such as its rate of growth, neighborhood, and the change in the role of a vertex over time. Independence of the temporal motif leads to mutually-exclusive motif instances by restricting each edge to participate in only one temporal motif instance. We use a set of the temporal motifs that are simple to compute but at the same time representative of temporal, structural, and functional properties of the network. We also define properties to measure the temporal evolution of the motifs, which informs the rate at which motifs are formed in the network. In contrast to previous work, no limit is put on the $\delta$ time window of the motif, but it can be restricted optionally. We provide algorithms to compute the independent temporal motif distribution of a given graph. Additionally, we also provide a new distributed implementation using the Apache Spark graph analytic framework.

The rest of the paper is organized as follows. Section \ref{sec:definition} lays out various definitions and section \ref{sec:tech_approach} presents our core approach. Section \ref{sec:experiments} shows our experimentation with synthetic and real-world temporal networks to summarize the temporal networks and measure their similarity. Section \ref{sec:conc} presents conclusions and future work.

\begin{table}[ht]
	\caption{Symbols and their descriptions}
	\begin{tabular}{ll}
		\textbf{Symbol} & \textbf{Description}                                     \\
		\hline
		$T$ & Temporal graph \\
		$G_i$ & $i^{th}$ window \\
		$t$ & Total number of windows\\
		$K$ & Set of Atomic Motifs \\
		$m_k$ & $k^{th}$ Atomic motif \\
		$m_{kl}$ & $l^{th}$ Temporal motif of $k^{th}$ atomic Motif\\
		$\mathcal{T}$ & Set of time-steps associated with motif edges \\
		$M$ & Motif instance \\
		$\hat{M}$ & ITeM instance \\
		$v_k$& Number of vertices in $k^{th}$ motif \\
		$\hat{V}_k$ & \# unique vertices in ITeM instances of $k^{th}$ motif\\
		$I$ & Set of Importance values for each window \\
		$I_i$ & Importance of $i^{th}$ window \\
		$\digamma $     & Temporal motif distribution for a given $T$ \\
		$d$ & \textit{Order} of a motif\\
		$o$ & \textit{Orbit} of a motif\\
	\end{tabular}
	\label{tab:freqconcept_table}	
\end{table}

\section{Definitions}\label{sec:definition} 
We present the ITeM-based approach to characterize a temporal network. In the following sections, we present definitions and algorithms used by ITeM to model a temporal network. We also review the Maximum Independent Set (MIS) problem, which is a subproblem of the proposed algorithm. MIS has been proved to be an NP-complete problem, and we present a heuristic-based approach to finding the lower bound on the ITeM frequency \cite{luby1986simple}. We also outline a sampling method to estimate the true frequency of a temporal motif in the network. The sampling approach is based on the \textit{importance} of the sampled network \cite{liu2018sampling}. 

A temporal graph is a specialization of a static graph, where each edge of the static graph appears at a time unit such as second, day, year, etc. Various representations of temporal graphs that are useful in different scenarios are proposed \cite{masuda2016guidance}. We use a window-based representation, where each window corresponds to a temporal sub-graph between two timestamps. 

\begin{definition}{\textbf{Temporal Graph:}}\label{def:tempgraph}
	A temporal graph \gls{T} is an ordered sequence of graphs $T=G_1,\dots,G_t$, indexed by a window id $i=1,\dots,t$. We define $\gls{Gi}=(V_i,E_i)$, where $V_i$ and $E_i$ denote the vertex and edge sets, respectively, in the window $i$, arriving since the window $i-1$. We say the temporal graph $T$ is on vertex set $V_T=V_1 \cup \dots \cup V_t$ and edge set $E_T=E_1 \cup \dots \cup E_t$. 
\end{definition}

This definition allows for the representation of a large graph with a single window. It is useful for datasets that are small in size and cover a small period of time.


\subsection{Atomic Motif}

\begin{figure}
	\centering
	\includegraphics[width=0.45\textwidth,height=80pt]{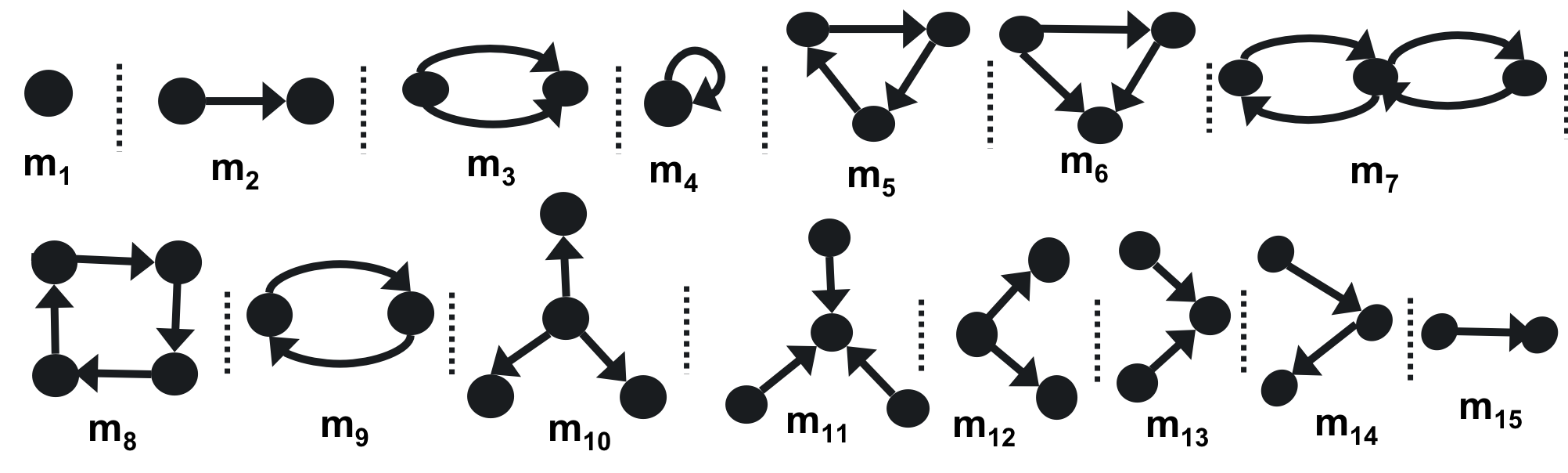}
	\vspace{-3mm}
	\caption{Atomic Motifs}
	\label{fig:atomicmotifs}
\end{figure}

Atomic motifs are small subgraphs that serve as interesting indicators for complex networks. They can reveal patterns of association among entities in the network. Figure \ref{fig:atomicmotifs} shows a library of atomic motifs used in the current work. Lower-order motifs such as isolated vertex (order d=1), self-loop (d=1), and isolated edge (d=2) are examples of \textit{fringe} motifs as they have less (sometimes zero) connectivity to the rest of the network. Whereas, higher-order motifs such as wedge (d=3), triangle (d=3), and square (d=4) are an example of \textit{core} motifs, which have been found to constitute a major fraction of real-world graphs. Our experimentation shows that the relative frequencies of \textit{fringe} and \textit{core} motifs in a temporal network can be used to compute graph similarity.

We can define atomic motifs of any number of vertices and edges, but the larger motifs are more difficult to search for in a network due to the intractability of the subgraph isomorphism, leading to an exponential increase in the runtime. Previous work shows that the computational cost of motif counting increases exponentially with k in $\mathcal{O}(|V|)^k$ \cite{shervashidze2009efficient}. Conversely, smaller atomic motifs are easier to find and yield better dividends in terms of modeling temporal and structural characteristics of the graph. 

We limit our motif library to 4-order motifs. The selection of d-order motifs to include in the search library has been influenced by previous research in this area, functional interpretation of the motifs in real-world domains, and computational pragmatism. In addition to the higher-order motifs  (d $>$ 2), we also make use of a few \textit{fringe} motifs that provide insight about a complex network that is not captured by such higher-order motifs. \textit{$m_1$} and \textit{$m_2$} correspond to isolated vertices and isolated edges in the network that are not part of any higher-order motif. An abundance of such motifs is a clear indicator of a sparse, disconnected state of the network and is important to model some domains, such as power-grids \cite{cuadra2015critical}. Similarly, \textit{$m_3$} and \textit{$m_4$} correspond to self-loop and multi-edges between the same set of entities. Frequencies of such motifs show important functional properties of the network and can be used to convert it into a smaller weighted network, where the self-loops and the multi-edges are converted into vertex and edge weights, respectively. At the same time, they also contribute to the combinatorial explosion of the higher-order motifs. The current set of motifs also allows us to analyze multiple domains without mining important subgraphs specific to that domain. While such subgraphs may better represent the domain, they require time and data to discover and would need to be limited in size to avoid the search complexity. We focus on using ITeM distribution for various downstream graph applications such as summarization, generation, and classification. 

\begin{figure}[h]
	\centering
	\includegraphics[width=0.45\textwidth,height=130pt]{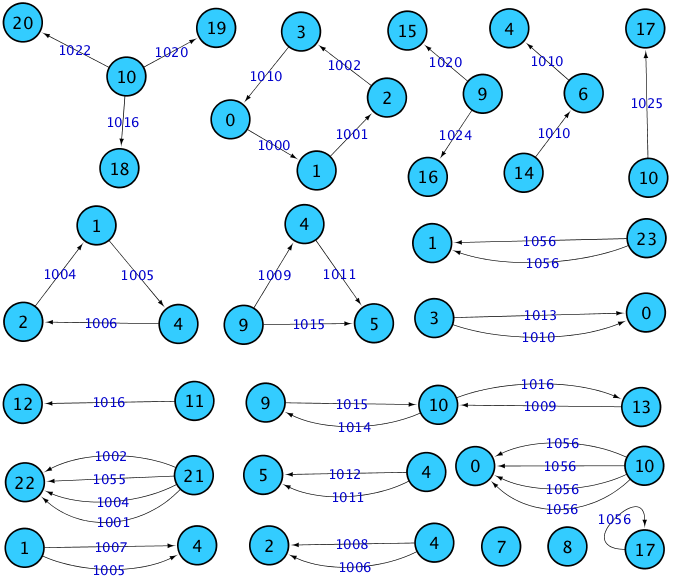}
	\caption{ITeMs for Example Input Graph}
	\vspace{-.6cm}
	\label{fig:testDifficutlGraph}
\end{figure}

Dyads and triads are the most used motifs to model complex networks. Larger acyclic and dense patterns do not uniquely explain different phases of temporal diffusion, whereas both the triads and linear chains do a better job \cite{sarkar2019understanding}. Motifs such as \textit{$m_5$} and \textit{$m_6$} are examples of feed back and feed forward loops \cite{benson2016higher} and are fundamental to understanding transcriptional regulation networks \cite{mangan2003coherent}, social networks, and biological systems. Adversarial activities exhibit patterns such as \textit{$m_7$} and \textit{$m_9$} among groups of adversaries, representing interactions such as communications and procurements \cite{cottam2018multi}. \textit{$m_7$} is also found on or around structural hubs in brain networks \cite{honey2007network}. The simple 4-cycle motif \textit{$m_8$} is an easy to find and informative structure.  Star motifs \textit{$m_{10}$} and \textit{$m_{11}$} are ubiquitous in many social networks. Two-hop paths such as \textit{$m_{12}$}, \textit{$m_{13}$}, and \textit{$m_{14}$} are essential to understanding air-traffic patterns \cite{benson2016higher} and procurement patterns \cite{cottam2018multi}. Directed wedges such as \textit{$m_{12}$} and  \textit{$m_{13}$} are also fundamental building blocks of bipartite graphs, which by definition do not show any triad or 4-cycle motifs. We also define \textit{residual edge motif} \textit{$m_{15}$}, which is a single edge, 2-vertex motif that represents instances of the interactions that are not discovered as part of any higher-order motif pattern and can be generated using a randomized network. The \textit{residual edge motif} \textit{$m_{15}$} differs from \textit{isolated edge motif} \textit{$m_{2}$} because \textit{$m_{15}$} represents the leftover edges in the graph at the end of the search order, whereas \textit{$m_{1}$} represents isolated edges found at the start of the analysis. As shown in Figure \ref{fig:testDifficutlGraph}, the edge $<11,12,1016>$ is an example of \textit{$m_{1}$}, and the edge $<10,17,1025>$ is a \textit{residual edge}.

\subsection{Temporal Motif}\label{sub:tm}

\begin{definition}{\textbf{Temporal Motif:}}
	A Temporal Motif $\mathcal{M}_t = (V, E, \mathcal{T})$ is a connected graph where:
	\begin{itemize}
		\item $V$ is a set of vertices of the motif.
		\item $E$ is a set of edges e $\in$ E, e:$(u,v, t), u \in V, v \in V, t \in \mathcal{T}$  where $\mathcal{T}$  is a set of time steps associated with motif edges.
		\item Edges have a temporal ordering such that for an edge $e_1$:$(u_1,v_1, t_1)$ and $e_2$:$(u_2,v_2, t_2)$ if $t_1 < t_2$ then $e_1$ arrives before $e_2$.
	\end{itemize} 
\end{definition}



\begin{figure}[h]
	\centering
	\includegraphics[width=0.5\textwidth,height=0.5\textheight]{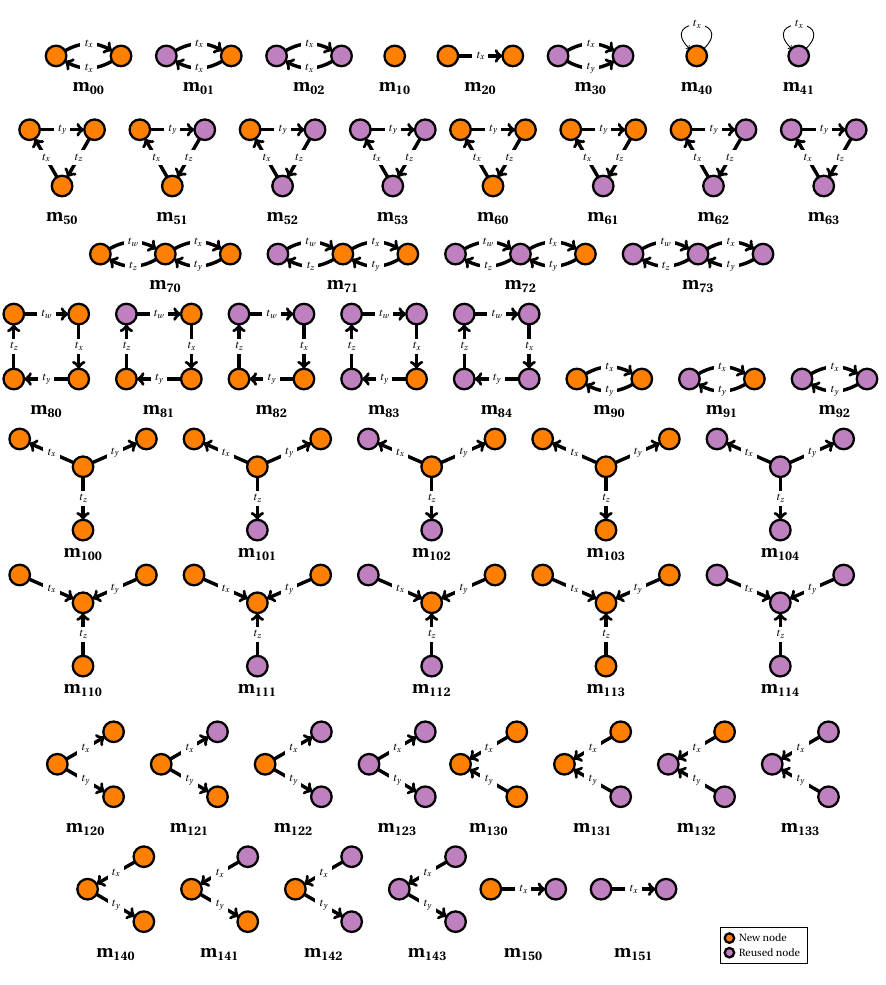}
	\vspace{-.6cm}
	\caption{Temporal Motifs}
	\label{fig:atomictemporalmotifs3}
\end{figure}

A Temporal Motif is a specialization of the atomic motif, where every interaction between two vertices occurs at a specific time-step.  The time-step $t_e$ of an edge $e$ defines a temporal ordering of the edge within the temporal motif $\mathcal{M}_t$. However, it does not correspond to the actual time of the interaction in the temporal graph. Using this definition, we extend the atomic motif to model its temporal evolution in terms of size and structure. Characterization of the temporal network using a set of static motifs can be misleading and inaccurate because the static motifs fail to capture the temporal properties of the network, such as the scale at which transactions occur \cite{benson2016higher}, burstiness of the transactions, and temporal dependency among the set of transactions. Additionally, many temporal systems are characterized as a dense \textit{multi-graph}, where a pair of entities share many temporal transactions as the network evolves. This poses additional combinatorial complexity challenges beyond discovering structural motifs in the network. Figure \ref{fig:atomictemporalmotifs3} shows a set of temporal motifs used in this work.

\begin{figure}[h]
	\centering
	\includegraphics[width=0.45\textwidth,height=130pt]{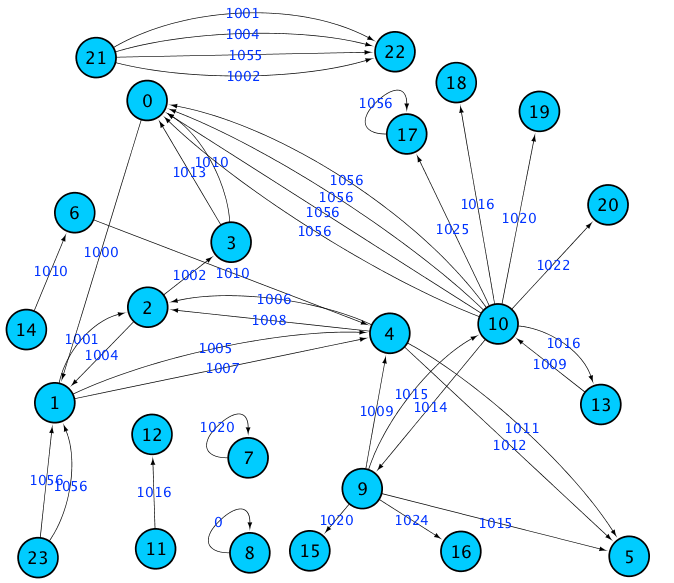}
	\caption{Example Input Graph}
	\label{fig:testDifficutlGraphNonMotif}
\end{figure}

\subsection{Independent Temporal Motif (ITeM)}\label{subsec:itm}  
Schreiber and Schwobbermeyer (\citeyear{schreiber2005frequency})  describe three different ways to measure the frequency of any pattern in a graph. They categorize them as \(F1\) , \(F2\), and \(F3\) \textit{concepts}. In the context of motif computation, \(F1\) includes every occurrence of a motif instance without any restriction, such as reusing a vertex or an edge while computing the frequency of motif instances. Paranjape et al. (\citeyear{paranjape2017motifs}) use this definition to compute overlapping $\delta$-motif frequencies. \(F2\) and \(F3\) concepts put restrictions on the reuse of a vertex or edge. \(F2\) is an edge-disjoint concept and does not allow the reuse of an edge in more than one instance of the motif. Similarly, \(F3\) is more restrictive as it is a vertex and edge-disjoint concept and does not allow reuse of any vertex and edge in more than one instance of the motif. 

A major contribution of our work is the ITeM, which is an edge-disjoint temporal motif such that no two motif instances share any edge between them. It is different than the temporal network modeling approaches mentioned in the related work, which use overlapping motif instances where some instances of a motif can share any number of edges. Overlapping motif instances can be used to model a network where it is common to have nodes and edges participate in multiple functional processes such as biological networks \cite{chen2006nemofinder} but fails to model a network where each edge represents one transaction between two entities such as a communication network. Independent motif instances capture a more accurate state of the network as no two transactions are part of any two motifs. Additionally, for a temporal network, ITeM can be used to model the rate at which the network grows as it distinguishes between adding a transaction using new nodes to the network and reusing them for multiple later transactions. Overlapping instances fail to capture this phenomenon as they compute all the isomorphic instances of a motif type. This restriction also poses a greater complexity issue as finding temporal motifs is proved to be an NP-Complete problem \cite{liu2018sampling}. In the following sub-sections, we define some key concepts used by ITeM to model a temporal network.

\subsubsection{\textbf{Vertex Birth-Time}}
We define the \textit{birth-time} of a vertex in the temporal network as the time of the first transaction involving the vertex. The \textit{birth} of a vertex increases the network size by one vertex. For the rest of the life of the network, that entity is treated as \textit{reused} and it never increases the network population. 
\subsubsection{\textbf{Structural Contribution}} Structural Contribution of an ITeM instance is a measure of the growth in the graph size as a result of adding the instance. The Structural Contribution of an independent temporal motif in terms of the number of edges is always equal to the number of temporal edges in the temporal motif. Figure \ref{fig:atomictemporalmotifs3} shows a set of temporal motifs and their \textit{structural contributions}. As shown in Figure \ref{fig:atomictemporalmotifs3}, every instance of $m_{62}$ adds three new temporal edges to an existing network. The \textit{structural contribution} in terms of the number of vertices is impossible to measure using static atomic motifs because an atomic motif instance fails to distinguish between the introduction of a new vertex to the network and reusing an existing vertex. Temporal motifs are required to encode this information to model the size and structure of the graph as it evolves. As shown in Figure \ref{fig:atomictemporalmotifs3}, every instance of the temporal motif $m_{62}$ adds only one new vertex to an existing network. Whereas, every instance of the temporal motif $m_{60}$ adds three new vertices to the temporal network.
\subsubsection{\textbf{Motif Orbit}}
An orbit of a motif is defined as distinct positions in which a vertex can appear within the motif. An $o-orbit$ motif has $o$ distinct positions. The orbit of a vertex in a motif encapsulates its functional role in the motif. As shown in Figure \ref{fig:atomicmotifs}, $m_5$ has just one orbit but $m_6$ has three different orbits. Similarity, star motifs $m_{10}$ and $m_{11}$  have two orbits each. A combination of \textit{structural contribution} and a change in the orbit of vertices allow us to model the evolution of a network without measuring the frequency of every \textit{automorphic} instance. Graph automorphism is a measure of the symmetry of a structure. It is defined as a mapping from the vertices of a given graph $T$ to itself.

\subsubsection{Independence:} We also define \textit{Independence} of a temporal motif as a measure of its uniqueness in a given temporal graph. The \textit{independence} can be measured for temporal motifs, temporal edges, or vertices of the temporal graph. The edge-disjoint concept defined above leads to maximal independent temporal edges because every edge has a bijection to the set of independent temporal motifs. We define the independence of a temporal motif and a vertex as follows:

\begin{definition}{Motif Independence:}
	For a given temporal motif $m_k$, the independence of the motif is defined as a ratio of the number of ITeM instances to the number of overlapping motif instances.
	\[
	DM_k= 
	\begin{cases}
	\frac{|\hat{M}_k|}{|M_k|},& \text{if } |M_k|\geq 0\\
	0,              & \text{otherwise}
	\end{cases}
	\] 
	where $|\hat{M}_k|$ is the total number of ITeM instances, and $|M_k|$ is the total number of motif instances ($|M_k| \geq |\hat{M}_k|$). 
\end{definition} 
This frequency-based metric identifies unique temporal motifs in the graph. Highly independent motifs exhibit the lower average cost of finding isomorphic combinatorial instances because of their uniqueness. 
\begin{definition}{Vertex Independence:}
	For a given temporal motif $m_k$, independence of the involved vertices is defined as a ratio of the number of unique vertices in ITeM instances to the maximum number of vertices possible in those instances. 
	\[
	DV_k= 
	\begin{cases}
	\frac{|\hat{V}_k|}{|M_k * v_k|},& \text{if } |M_k|\geq 0\\
	0,              & \text{otherwise}
	\end{cases}
	\]
	where $|\hat{V}_k|$ is the number of unique vertices in the ITeM instances of the $k^{th}$ motif, $|M_k|$ is the total number of motif instances, and $v_k$ is the number of vertices in the $k^{th}$ motif.
\end{definition} 
Temporal motifs with high vertex independence lead to high \textit{structural contribution}, whereas low vertex independence leads to co-located independent temporal motifs with a higher number of shared vertices among them.

\section{Approach} \label{sec:tech_approach} 
In this section we present an exact algorithm to count ITeM frequency. We also present an approximate algorithm using Importance sampling. 
\subsection{Exact algorithm to count ITeM frequency}

\begin{algorithm}[h]
	\SetAlgoLined
	\LinesNumbered 
	\KwData{$T \colon$ Temporal Graph}
	\KwData{$K \colon$ Set of Atomic Motifs}
	\KwResult{$M_n \colon$ Independent motif instances}
	
	\ForEach{ $m_k \in$ $K$}{
		$M  \leftarrow $ getMotifInstances$(m_k, T)$\\
		$M_n \leftarrow M_n \cup $ getITeM$(M)$\\
	}
	\textbf{return} $M_{n}$
	\caption{ITeM($T,K$)}
	\label{algo:ITeMMain}
\end{algorithm}
\begin{algorithm}[h]
	\SetAlgoLined
	\LinesNumbered 
	\KwData{$M \colon$ All motif instances}
	\KwResult{$M_n \colon$ Independent motif instances}
	\footnotesize\ttfamily\textcolor{blue}{\tcc{Create a mapping $EM$ between an edge $e$ and all associated motif instances. $\mathcal{L}(i)$ is the string representation of a motif instance \textit{i}.}}
	\ForEach{i $\in$ $M$}{
		\ForEach{e $\in$ i}{$EM(e) \leftarrow EM(e) \cup \mathcal{L}(i) $}
	}
	
	\footnotesize\ttfamily\textcolor{blue}{\tcc{For every motif instance label $i$, create a vertex in the overlap graph. }}
	\ForEach{ $i \in$ M}{
		$H_v \leftarrow H_v \cup \mathcal{L}(i)$
	}
	\footnotesize\ttfamily\textcolor{blue}{\tcc{Create an edge in the overlap graph, between every motif-instance label pair $(l_r, l_{r+1})$ that share an edge $e$ in the input graph. }}
	\ForEach{e $\in$ EM}{
		\ForEach{ ($l_r, l_{r+1}) \in$ EM(e)}{
			$H_e \leftarrow  H_e \cup (l_r, l_{r+1})$
	}}
	\footnotesize\ttfamily\textcolor{blue}{\tcc{Create the motif overlap graph}}
	$H \leftarrow G(H_v,H_e)$\\
	\footnotesize\ttfamily\textcolor{blue}{\tcc{Find non-overlapping temporal instances}}
	$M_n \leftarrow MaxIndSet(H)$
	
	\textbf{return} $M_{n}$
	\caption{getITeM($M$)}
	\label{algo:nonOverlappingM}
\end{algorithm}

\begin{algorithm}
	\SetAlgoLined
	\LinesNumbered 
	\KwData{$H \colon$ An undirected abstract graph}
	\KwResult{$I \colon$ Maximum Independent set of vertices}
	\footnotesize\ttfamily\textcolor{blue}{\tcc{Set every vertex in its own Independent Set}}
	\ForEach{$v \in H_v$}{$I_v$ = $\mathcal{L}(v)$}
	\Repeat{$I_v$ does not change}{
		send $ I_v \colon v \in H_v$ to every $u \in Neighbor(v)$\\
		receive $ I_u$ for every $u \in Neighbor(v)$\\
		update $I_v$ by the lowest $ I_u$ received
	}
	\footnotesize\ttfamily\textcolor{blue}{\tcc{Get Independent Set as unique values of $I_v$}}
	\textbf{return} $unique(I) $
	\caption{MaxIndSet($H$)}
	\label{algo:MaxIndSet}
\end{algorithm}

Finding matches to temporal motifs is proved to be an NP-Complete problem \cite{liu2018sampling}. We use Luby's Algorithm \cite{luby1986simple} to discover ITeMs which provides a lower bound on the ITeM frequency. 

Algorithms \ref{algo:ITeMMain}, \ref{algo:nonOverlappingM}, and \ref{algo:MaxIndSet} present the pseudocode to find independent temporal motif instances in a given temporal graph.  Algorithm \ref{algo:ITeMMain} inputs a temporal graph and a set of atomic motif types to discover as shown in Figures \ref{fig:testDifficutlGraphNonMotif} and \ref{fig:atomicmotifs} respectively . Line 1 discovers all overlapping motif instances of a given motif type $m_k$. We generate all the temporal motif types corresponding to $m_k$  (Figure \ref{fig:atomictemporalmotifs3}) and use GraphFrame \cite{dave2016graphframes} to discover the overlapping temporal motif instances. Overlapping motif discovery is a run-time bottleneck and GraphFrame provides optimized motif discovery using graph-aware dynamic programming algorithms. It also provides a simple Domain-Specific Language (DSL) to express all the temporal motifs.  Algorithm \ref{algo:nonOverlappingM} inputs a set of overlapping temporal motif instances discovered in Line 1 and returns ITeM instances (as shown in Figure \ref{fig:testDifficutlGraph}).  We use temporal ordering of the edges to define $\mathcal{L}(m)$, a lexical representation of the motif instance. It is used as a vertex label to construct a motif overlap graph $H$. The motif overlap graph $H$ is an abstract graph that represents clusters of motif instances sharing at least one edge in the input graph $T$ as defined in Definition \ref{def:tempgraph}. Lines 2-6 map an edge and its associated set of motif instances. Lines 8-10 create a set of vertices $H_v$ in the abstract graph. Lines 12-16 construct an edge-list $H_e$ using all the motifs that share a temporal edge in the input graph. $H_e$ is constructed by creating an edge in the abstract graph $H$ for every shared edge in the input graph $T$. $H_e$  and $H_v$ are used to construct the abstract graph $H$ on Line 18. The final result is computed using Algorithm \ref{algo:MaxIndSet} on Line 20, which uses a distributed MIS implementation to compute the ITeM instances. The ITeM instances represent a set of edge-disjoint motif instances in the input graph. 

Algorithm \ref{algo:MaxIndSet} presents the pseudocode of a distributed implementation of the MIS algorithm. We use Pregel API, available in Apache Spark, to implement Luby's Algorithm \cite{luby1986simple}. We initialize all vertices in their own independent set as shown in lines 2-4. At lines 5-9 of Algorithm \ref{algo:MaxIndSet}, each vertex exchanges messages with its neighbors and updates its independent set value based on the minimum values received from all neighbors. This process stops when no vertex in the graph changes its independent set.

\subsection{Approximate algorithm to count ITeM frequency}
Our approach includes three major algorithmic components: searching for overlapping temporal atomic motifs, finding independent temporal motifs, and computing information content and temporal evolution of such motifs.  Out of the three components, finding independent temporal motifs is an NP-Complete problem, and we use a heuristic to find a lower bound of the actual count. As explained in the previous section, we construct a motif overlap graph where every vertex is a motif instance and an edge between two vertices exists if the corresponding motif instances share an edge in the original temporal graph $T$. This abstract formulation may lead to a highly-cliqued abstract graph, which is a characteristic of various real-world domains, such as a social network. A highly-cliqued abstract graph leads to excessive message-passing in the distributed computing environment. To address this, we use an \textit{importance} based sampling approach to approximate the \(F2\) motif frequency computation.

Importance sampling for motifs is presented by Liu et al. (\citeyear{liu2018sampling}). It is based on the assumption that each distribution has some interesting or important regions and the samples drawn from those regions must be normalized to get an unbiased estimate \cite{mcbook}. Window-based importance sampling \cite{owen2013monte} splits the time series dataset into multiple temporal windows and performs exact computation on each window. We create window graphs with equal temporal window size, each with a different number of edges within the window. Each window is assigned an \textit{importance}, based on the fraction of all the edges present in the window. It is used to normalize the computed metric across all randomly-selected windows. The normalization reduces the overall variance for real-world domains that do not show a \textit{burst}. The current approximation approach does not model such anomalies in the ITeM distribution but allows us to model the evolution of a network as shown in  section \ref{sec:experiments}. Future work will address this challenge using an importance decay approach that gives more importance to recent windows. We compute the distribution of all temporal motifs present in the window graph. At the end of all the windows, we compute the weighted average of all the distributions, which gives a lower bound estimate of the distribution that can maintain a relative error tolerance of 5\% in the count \cite{liu2018sampling}.


For a given temporal graph $T$ with $t$ windows, the importance vector \gls{ImpAll} is an ordered sequence of \textit{window importance} \gls{ImpI}: $\mathcal{I} =  <I_{1}, I_{2},..., I_{t-1}, I_{t}>$
where the $I_i$ is defined as: $I_i = \frac{|E_i|}{|E_T|}$
where $E_i$ is the number of edges in a window \textit{i} and $E_T$ is the total number of edges in the temporal graph.
For a given motif $m_k$, the expected motif frequency \gls{Fm} in the temporal graph can be computed from the exact frequency $\Delta_{ki}$ of the motif in the $i^{th}$ window with importance $I_i$ as:
\[ f_{ki} =  \frac{\Delta_{ki}}{I_i}   \quad\text{and}\quad  \digamma_{k} = \frac{1}{t}\sum_{i=1}^{t} f_{ki} \]
We also define a random variable $X_i \in \{0,1\}$ that selects a specific window in the entire population. The expected frequency $\digamma_{k} $  is computed as :
\[ \digamma_{k} = \frac{1}{t_x}\sum_{i=1}^{t_x} X_i * f_{ki}\] where $t_x$ is the number of windows selected ($X_i = 1$) for the ITeM disovery.The ITeM distribution \gls{F} for a given temporal graph is the distribution of all such temporal motifs over the window population. $\digamma = < \digamma_{1},  \digamma_{2}, ...,  \digamma_{K} >$ where $|K|$ is the total number of motifs.



\section{Experiments} \label{sec:experiments}
To evaluate the performance and scalability of our approach, we analyzed a rich set of synthetic and real-world temporal datasets. The experiment provides support for our following core contributions:
\begin{itemize}
	\item ITeMs are a novel way of capturing discerning temporal properties of a temporal network that cannot be measured using static and overlapping temporal motifs.
	\item ITeMs outperform the Stanford SNAP temporal motif algorithm (referred as $\delta$-Motif hereinafter)  and Dynamic Graphlet (DG)  \cite{hulovatyy2015exploring} in measuring the similarity of temporal graphs.
	\item Our approach is scalable and configurable to analyze a temporal network as one large graph or a sequence of windows using sampling.
\end{itemize} 

All the experiments are done on a cluster using Apache Spark 2.3.0 and GraphFrame 0.7.0. All the algorithms are implemented in Scala 2.11.8, and the source code is available at 
\textcolor{blue}{https://github.com/temporal-graphs/STM}.

\subsection{Results on Synthetic Networks}\label{sec:results_synth}
ITeMs can efficiently model the evolution of a temporal network using the properties defined in the section above. To present the accuracy of modeling temporal changes in the network using ITeMs, we generate a set of synthetic temporal graphs using a stochastic generation method and measure the change in the similarity as the networks evolve. We benchmark against $\delta$-Motif and DG and show that ITeMs are better at measuring the changes in the similarity as the networks evolve. For a given population size $|V|$ = 100, we create a temporal graph $G_0$ of one-day time duration, where every vertex creates an edge with a random target vertex with a low probability $p$ at every second. Then, we create variations of the base graph by stretching it one day at a time and perturbing timestamps using a Gaussian distribution with zero mean and 1/6 day as standard deviation. We create thirty such variations (${G_1, G_2,......G_{30}}$). For example, the time between edge arrivals in $G_{10}$ is 10 days longer than in $G_1$. All the graphs in the sequence have the same structure and only the edge timestamps vary. 

\begin{figure}[h]
	\centering
\includegraphics[width=0.5\textwidth,height=0.2\textheight]{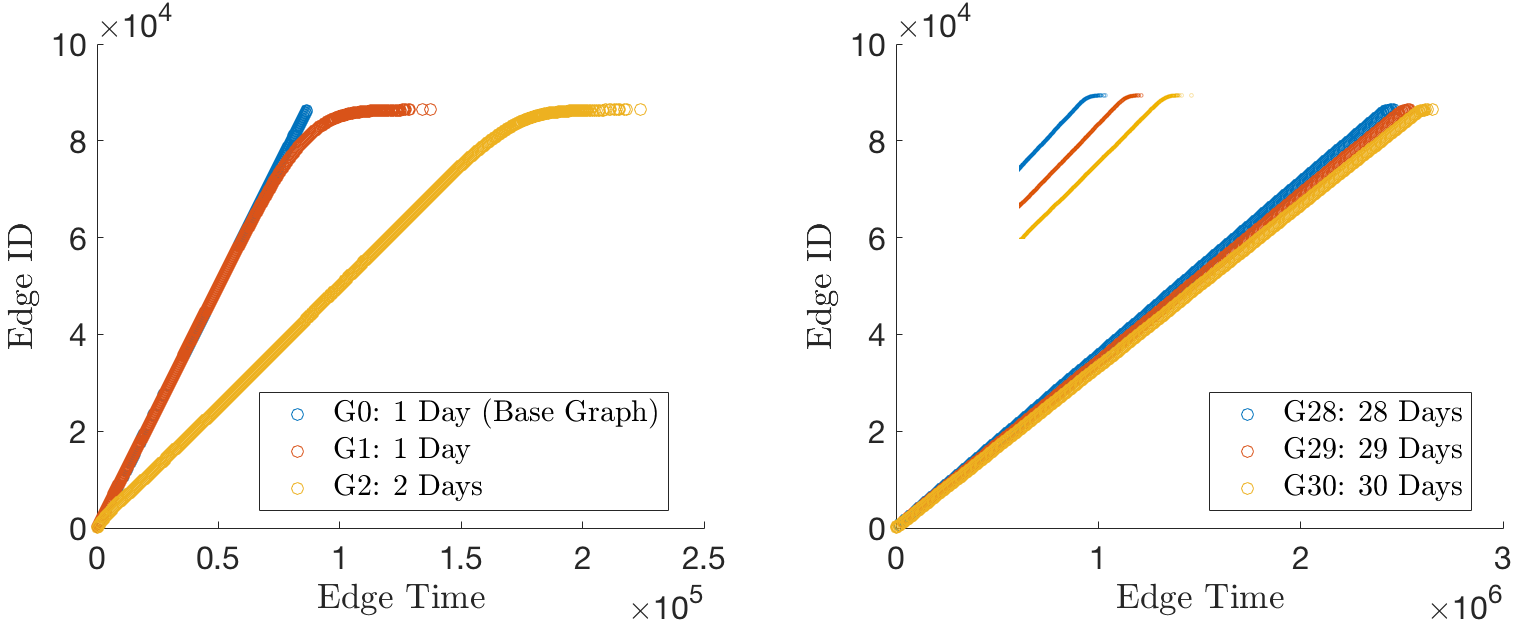}
\caption{ Synthetic Graphs}
\label{fig:synthG1G28}
\end{figure}

\begin{figure}[h]
	\centering
	\includegraphics[width=0.5\textwidth,height=0.2\textheight]{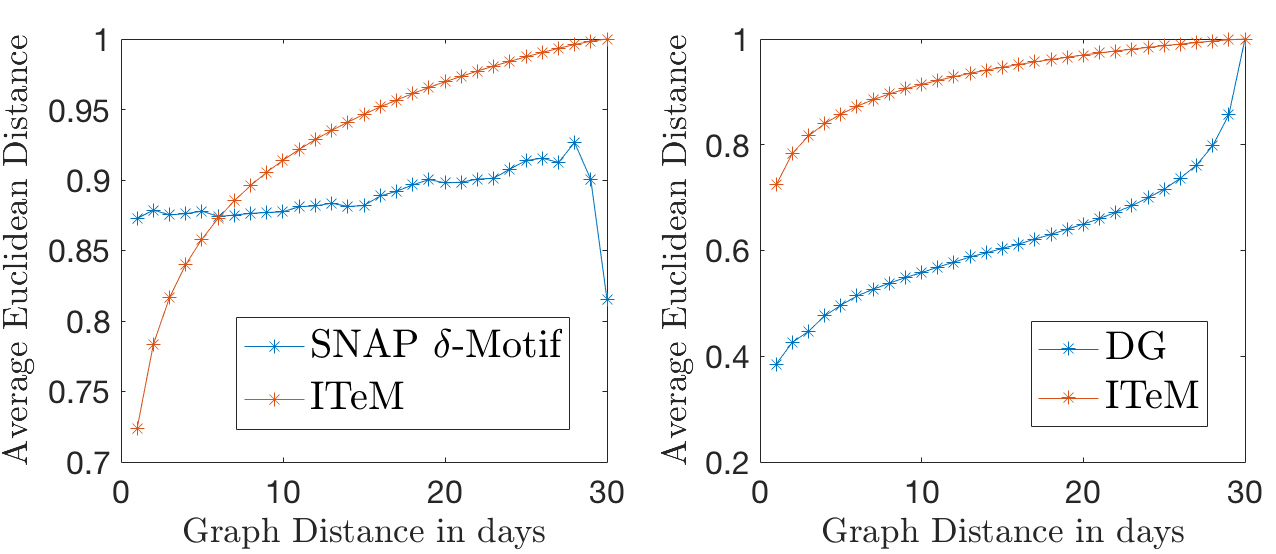}
	\vspace{-.3cm}
	\caption{Temporal Graph Similarity }
	\vspace{-.5cm}
	\label{fig:snapdgitem}
\end{figure}

Figure \ref{fig:synthG1G28} shows the rate of the addition of temporal edges to the graph. We also show a zoomed-in version (right) of $G_{28}, G_{29},$ and $G_{30}$ to visualize linearity in the temporal stretch as we increase the total time of the graph. We compute motif frequencies using both algorithms. Similarly, we also compute temporal, structural, and orbital features using our ITeM approach. For a given approach, we compute a feature matrix with 31 rows where each row corresponds to one synthetic network. Additionally, each row represents a fixed-column vector where the length of the row corresponds to total features computed by the tool. These feature vectors (i.e., embeddings) are used to measure the pairwise similarity of the temporal networks. Once we compute the pairwise similarities for all the networks, we aggregate them for the networks with same temporal stretch. This explains Figure \ref{fig:snapdgitem} where each entry (i,j) represents \textit{j} avg. Euclidean  distance for all the networks with the total duration \textit{i} days apart. The experiment is repeated for all the three approaches. 



Figure \ref{fig:snapdgitem} shows the change in normalized graph similarity as a function of the difference in the time duration of the synthetic graphs. A point (i,j) on the plot represents the average Euclidean distance j over all the graphs that are \textit{i} days apart. The $\delta$-Motif allows the use of arbitrarily large $\delta$ values (the limit on the time window spanned by motifs), and we use this feature to identify motifs without any temporal restriction on the time difference between any two motif edges. Figure \ref{fig:snapdgitem} (left) shows that the temporal-spatial-orbital features computed by ITeM outperform graph similarity accuracy using $\delta$-Motif features that are based only on motif counts. The $\delta$-Motif does not capture the temporal variations of discovered motif instances, whereas ITeM can successfully measure it as the graph is \textit{stretched} in time and the average $\delta$ time between edges and the time to form a motif increases. For maximum distant graphs such as $G_0$ and $G_{30}$, we observe an unexpected sharp change in the similarity using $\delta$-Motif. This requires a deeper analysis of the algorithm and the output generated by the tool. 

DG also characterizes a temporal network in terms of \textit{graphlet} count for the entire network and individual nodes. DG also provides a $\delta$ parameter to restrict time difference between two edges of the graphlet, but due to out-of-memory errors, we could not run it in the unbounded setup that was used in the previous experiment. To benchmark against DG, we used a $\delta$ restrictive mode of our algorithm with $\delta$ set to 600 seconds. 

Figure \ref{fig:snapdgitem} (right) shows the result comparing DG and ITeM. As shown in the Figure \ref{fig:synthG1G28}, the base graph shifts from a stochastic base model to a Gaussian distribution based temporal network, which explains the initial sharp increase in the graph distance measured by both algorithms. Both the approaches also show sub-linear trends afterward but only ITeM continues as the time difference between graphs increases. DG shows sudden exponential changes in the distance (or similarity) that do not correspond to the linear temporal evolution of the graphs as shown in Figure \ref{fig:synthG1G28} (right). Overall, both the approaches exhibit similar trends that show the importance of modeling temporal variations and orbital information of the graph, in addition to the frequency count.

\begin{table}
	\caption{Temporal Graphs Datasets}
	\vspace{.3cm}
	\label{table:allGraphs}
	\begin{tabular}{ r|c|c|c|c| }
		\multicolumn{1}{r}{}
		&  \multicolumn{1}{c}{$|V|$}
		& \multicolumn{1}{c}{$|E_{temporal}|$} 
		& \multicolumn{1}{c}{$|E_{static}|$}
		& \multicolumn{1}{c}{Time} \\
		\cline{2-5}
		CM & 1,899 & 59,835 & 20,296 & 193 days \\
		\cline{2-5}
		BA& 3,783 & 24,186 & 24,186  & 1,901 days \\
		\cline{2-5}
		EE & 986 & 332,334 & 24,929 & 803 days \\
		\cline{2-5}
		TT & 34,800 & 171,403 & 155,507 & 21 hours \\
		\cline{2-5}
		IA & 545,196 & 1,302,439 &  1,302,253  & 1,153 days \\
		\cline{2-5}
		HT & 304,691  & 563,069 &  522,618 & 7 days \\
		\cline{2-5}
		RH & 55,863 & 571,927 & 561,483 & 3 yrs 4 mos \\
		\cline{2-5}
		WT & 1,140,149 & 7,833,140 & 3,309,592 & 6 yrs 4 mos \\
		\cline{2-5}
	\end{tabular} 
\end{table}
\vspace{-.2cm}
\subsection{Results on Real-World Networks}\label{sec:results_real}
We analyze various real-world networks and measure the difference in their temporal evolution. The following list introduces all the datasets used for the experiments. Table \ref{table:allGraphs} describes their static and temporal scale. We generate tITeM distribution and use it for the measurement. We also use the change in the distribution over time to detect an event in the network.

\begin{figure}
	\centering
	\includegraphics[width=.5\textwidth,height=230pt]{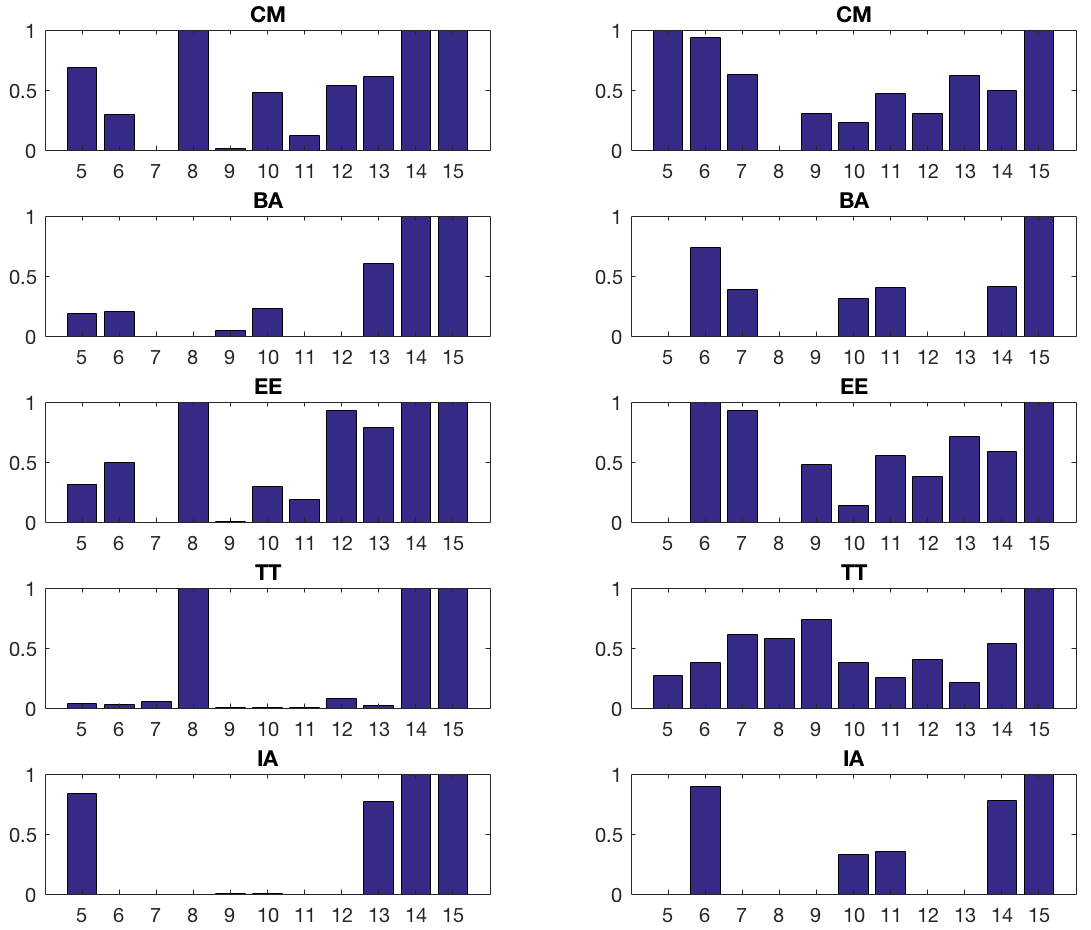}
	\vspace{-.5cm}
	\caption{Motif  and Vertex Independence of different datasets. x-axis represents motif-id and y-axis represents Motif Independence (left) and Vertex Independence (right)}
	\label{fig:motifIndresult}
\end{figure}

\begin{itemize}
	\item \textbf{CollegeMsg (CM)}: CollegeMsg \cite{panzarasa2009patterns} is comprised of private messages sent on an online social network at the University of California, Irvine. An edge (u, v, t) means that user u sent a private message to user v at time t.
	\item  \textbf{Bitcoin-Alpha (BA)}: Bitcoin-Alpha \cite{kumar2016edge} is a who-trusts-whom network of people who trade on \textit{Bitcoin Alpha} platform. An edge (\textit{u}, \textit{v}, \textit{t}) in the network exists if person \textit{u} gives a rating to person \textit{v} at time \textit{t}.
	\item \textbf{Email-EU (EE)}: Email-EU \cite{yin2017local} \cite{leskovec2007graph} is an anonymized network about all incoming and outgoing emails between members of a large European research institution. An edge (\textit{u}, \textit{v}, \textit{t}) in the network exists if person \textit{u} sent an email to person \textit{v} at time \textit{t}.
	\item \textbf{Tech-As-Topology (TT)}: Tech-As-Topology \cite{nr} is a temporal network of Autonomous Systems (AS) where an edge (\textit{u}, \textit{v}, \textit{t}) represents a link between AS \textit{u} and AS \textit{v} at time \textit{t}.
	\item \textbf{IA-Stackexch (IA)}: IA-Stackexch-User-Marks-Post is a bipartite Stack Overflow favorite network \cite{nr}. Nodes represent users and posts. An edge  (\textit{u}, \textit{v}, \textit{t}) denotes that a user \textit{u} has marked a post \textit{v} as a favorite at time \textit{t}.
	\item \textbf{Higgs Twitter: (HT)} Higgs dataset \cite{de2013anatomy} is an anonymized network that has information about messages posted on Twitter between the 1st and the 7th of July 2012 about the announcement of the discovery of Higgs boson particle. An edge (\textit{u}, \textit{v}, \textit{t}) represents a Twitter interaction between user \textit{u} and \textit{v} at time \textit{t}. An interaction can be a \textit{re-tweet}, \textit{mention}, or \textit{reply}.
	\item \textbf{Reddit Hyperlink (RH)}: Reddit hyperlink \cite{kumar2018community} represents the directed connections between two subreddits. An edge (\textit{u}, \textit{v}, \textit{t}) represents a hyperlink from subreddit \textit{u} to subreddit \textit{v} at time \textit{t}.
	\item \textbf{Wiki-talk (WT)}: Wiki-talk \cite{paranjape2017motifs} represents Wikipedia users editing each other's Talk page. A directed edge (u, v, t) means that user u edited user v's talk page at time t.
\end{itemize}	
\begin{figure}[h]
	\centering
	\includegraphics[width=.45\textwidth,height=70pt]{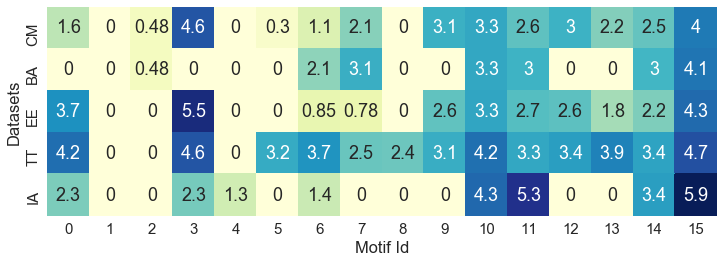}	
	\vspace{-.5cm}
	\caption{ITeM distribution (log10) of different datasets}
	\vspace{.1cm}
	\label{fig:itmresult}
\end{figure}
Figure \ref{fig:itmresult} shows the independent temporal motif distribution of different datasets.  Similarly, Figure \ref{fig:motifIndresult} shows motif independence and vertex independence for the datasets. These results give initial clues that similar domain networks such as \textbf{CM} and \textbf{EE} exhibit similar motif and vertex independence, whereas \textbf{BA} and \textbf{TT} have a different distribution. 
\begin{figure}[h]
	\centering
\includegraphics[width=.45\textwidth,height=180pt]{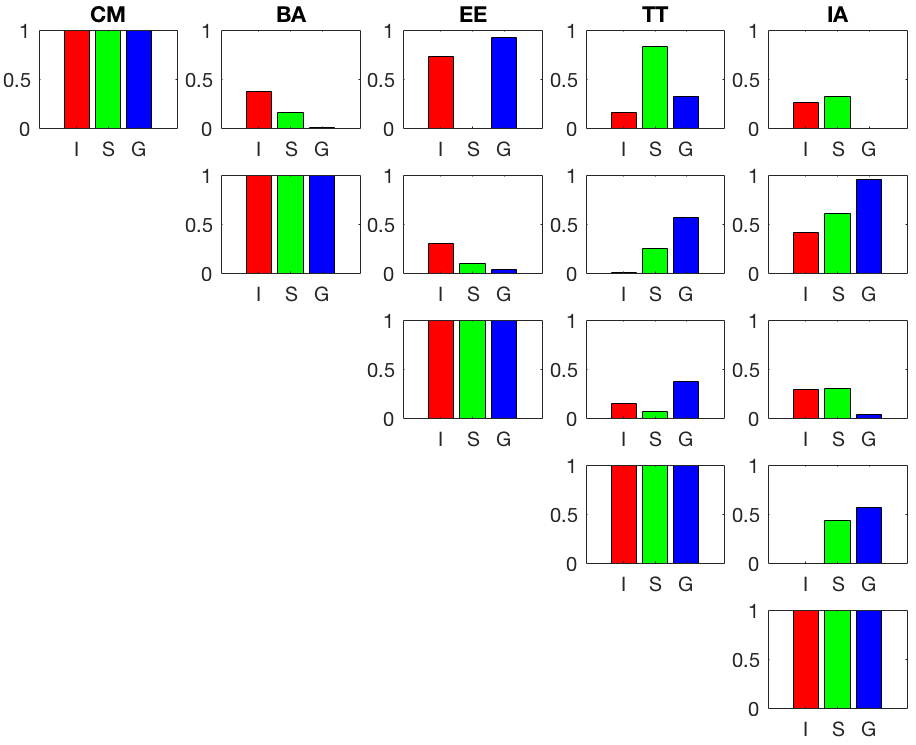}	
	\vspace{-.3cm}
	\caption{Real-world Graph Similarity (y-axis) using ITeM, SNAP, and DG (x-axis) for CM, BA, EE, TT, and IA}
	\label{fig:realworld_results}
\end{figure}
We also analyze real-world datasets using ITeM, SNAP $\delta$-Motif and DG. In the absence of any ground-truth, we observe the types of motifs and their frequencies discovered by the tools. We restrict our analysis to the motifs of maximum 4 vertices. ITeM can identify \textit{fringe} motifs such as isolated nodes, isolated edges, and self loops. DG does discover single edges but not the isolated nodes and self-loops. Both DG and SNAP $\delta$-Motif focus on connected networks. All three tools discover multi-edges in the network. 

We also compute network similarity across all network pairs, using Euclidean distance between the normalized frequency vectors. Figure \ref{fig:realworld_results} shows the similarity between each pair of real-world datasets using the three approaches. Both DG and ITeM identify CollegeMsg (\textbf{CM}) more similar to Email-EU (\textbf{EE}) than Bitcoin-Alpha (\textbf{BA}) and Tech-As-Technology (\textbf{TT}). We could not run DG in the unbounded setup so we restricted $\delta$ to 6000 seconds and that leads to single edge only motifs in the case of \textbf{BA} and \textbf{IA}. Similarly, SNAP $\delta$-Motif discovers few non-negative motif instances and we treat them as zero for the analysis. ITeM and $\delta$-Motif also generate a fixed size feature vector for a given input which makes it easier to use in downstream applications.

\begin{figure*}
	\includegraphics[width=\textwidth,height=140pt,]{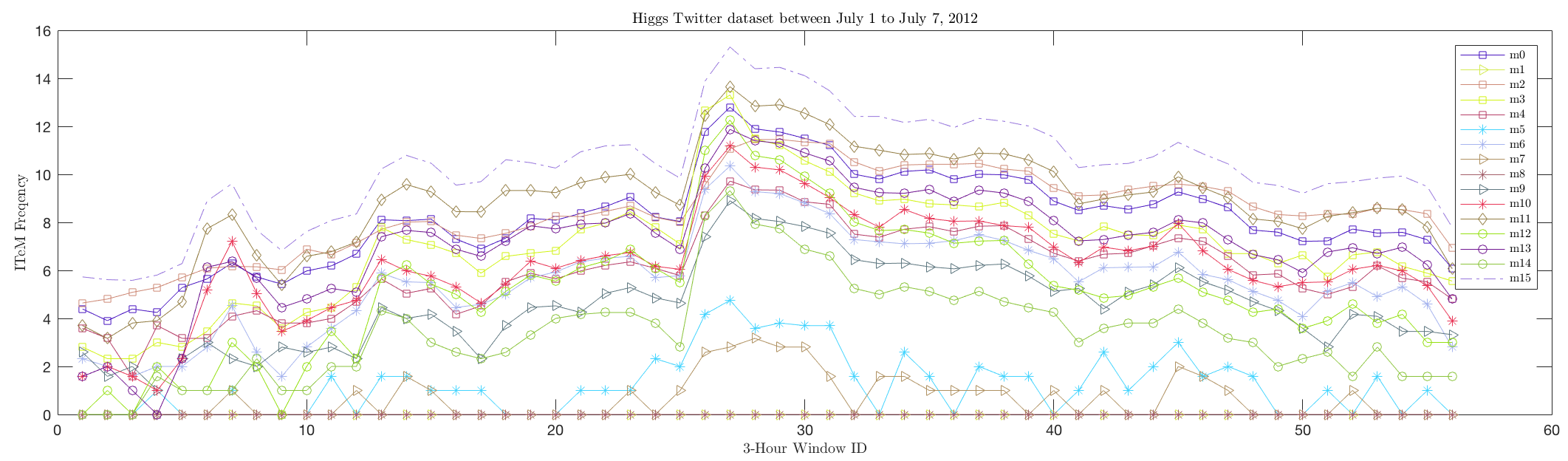}
	\vspace{-.7cm}
	\caption{ITeM frequency changes in the Higgs Twitter (HT) temporal network }
	\label{fig:higgsresult}
\end{figure*}

ITeM can also model the temporal evolution of a network using a sequence of temporal graphs, each with a given time window. We use the Higgs Twitter (\textbf{IT}) dataset and monitor 3-hour windows from July 1st to July 7th. Our approach iteratively analyzes each window and updates the temporal summary of the network as it progresses. This allows us to not only analyze a large graph using multiple smaller graphs but also to identify an anomalous event in the network and to understand how the behavior of vertices changes in the temporal network. Figure \ref{fig:higgsresult} shows a change in ITeM frequencies to reflect a burst event in the graph. The ITeM frequencies peak at the event on July 4th and then gradually return to a normal state. ITeM also provides more insight into the event than basic graph density-based measures. As shown in Figure \ref{fig:higgsresult}, the maximum increase is observed in the \textit{fringe} part of the network, such as self-loops, isolated edges, and residual edges. Similarly, a higher number of stars and wedges are also observed. These observations correspond to a network growth phenomenon where a burst of new interactions occurs in the network among newly-added entities. In the case of \textbf{HT}, this is explained by a higher number of Twitter users tweeting about the Higgs boson particle discovery for a short period of time. 

\begin{figure*}
	\includegraphics[width=\textwidth,height=140pt]{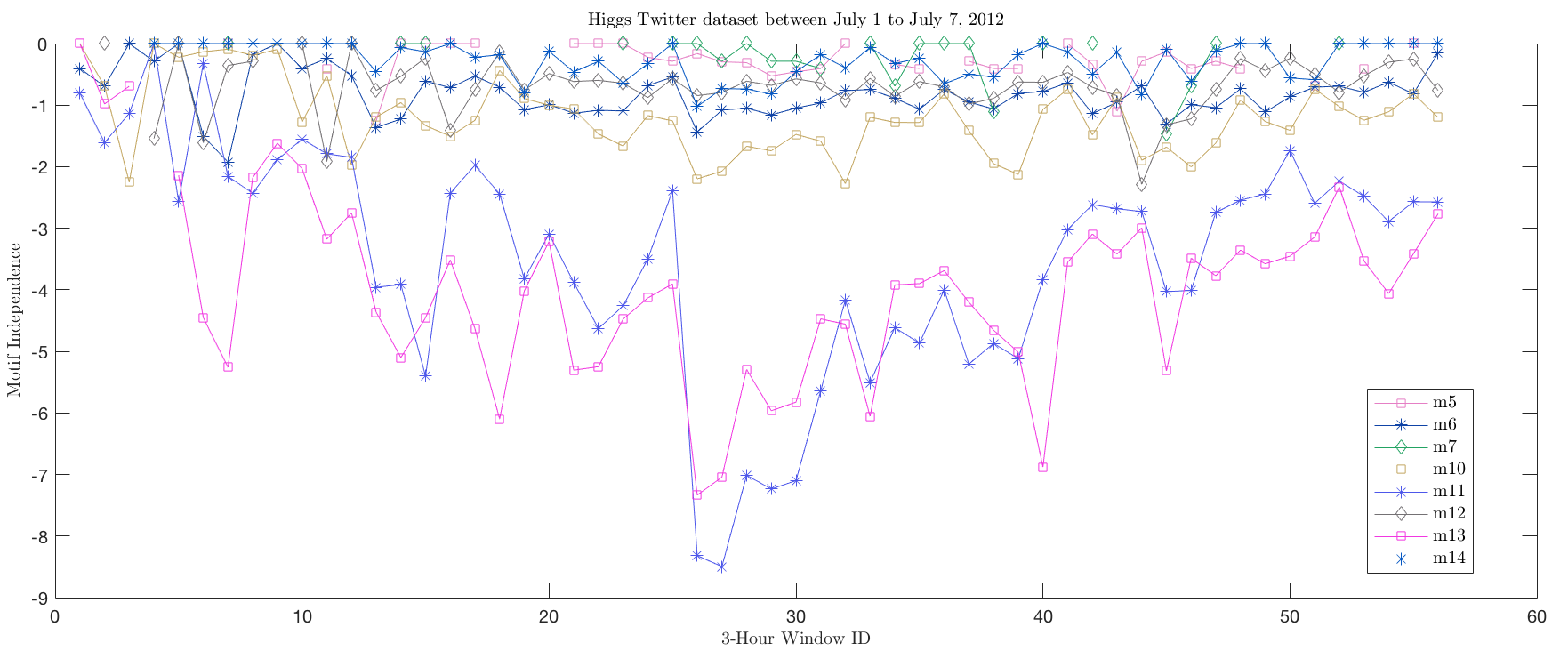}
	\vspace{-.7cm}
	\caption{ITeM Independence changes in the Higgs Twitter (HT) temporal network }
	\label{fig:higgsresultInd}
\end{figure*}

\begin{figure}[h]
	\includegraphics[width=.5\textwidth,height=110pt]{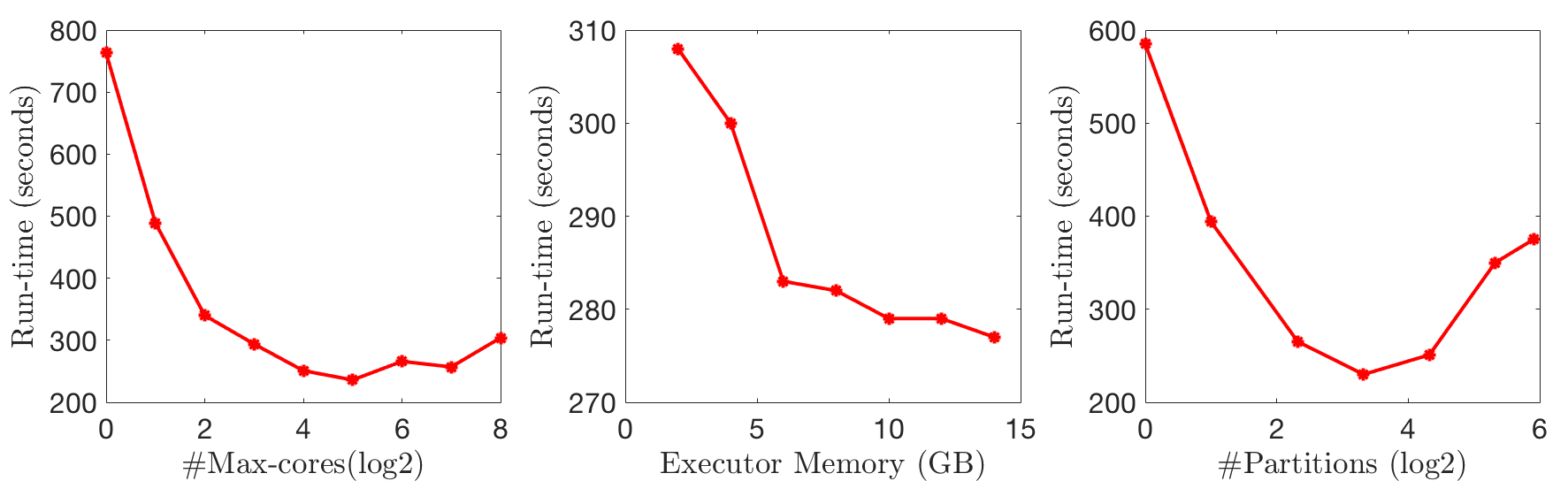}
	\vspace{-.6cm}
	\caption{ITeM runtime analysis on Email-EU (EE) dataset: Single Graph }
	\label{fig:emailScalability}
\end{figure}

\begin{figure}
	\includegraphics[width=.5\textwidth,height=110pt]{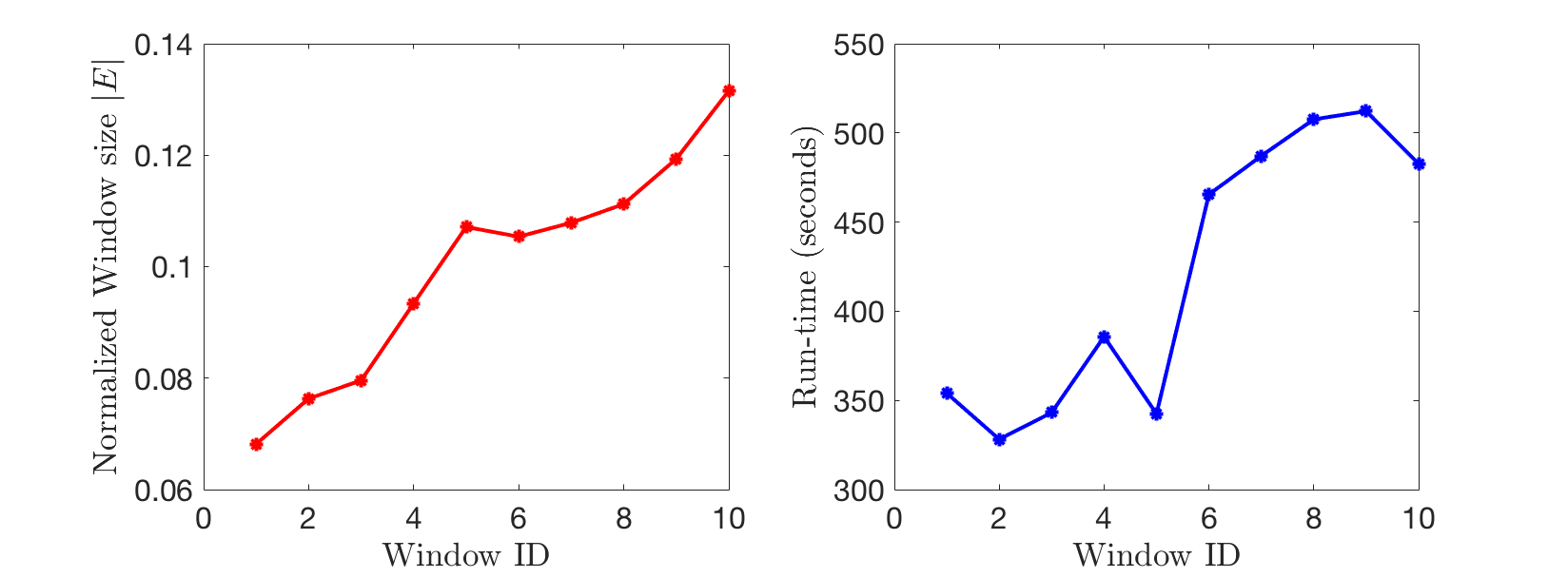}
	\vspace{-.6cm}
	\caption{ITeM runtime analysis on Reddit (RH) dataset: Sequence of Temporal Graphs}
	\label{fig:redditruntime}
\end{figure}

\begin{figure*}
	\includegraphics[width=\textwidth,height=140pt,]{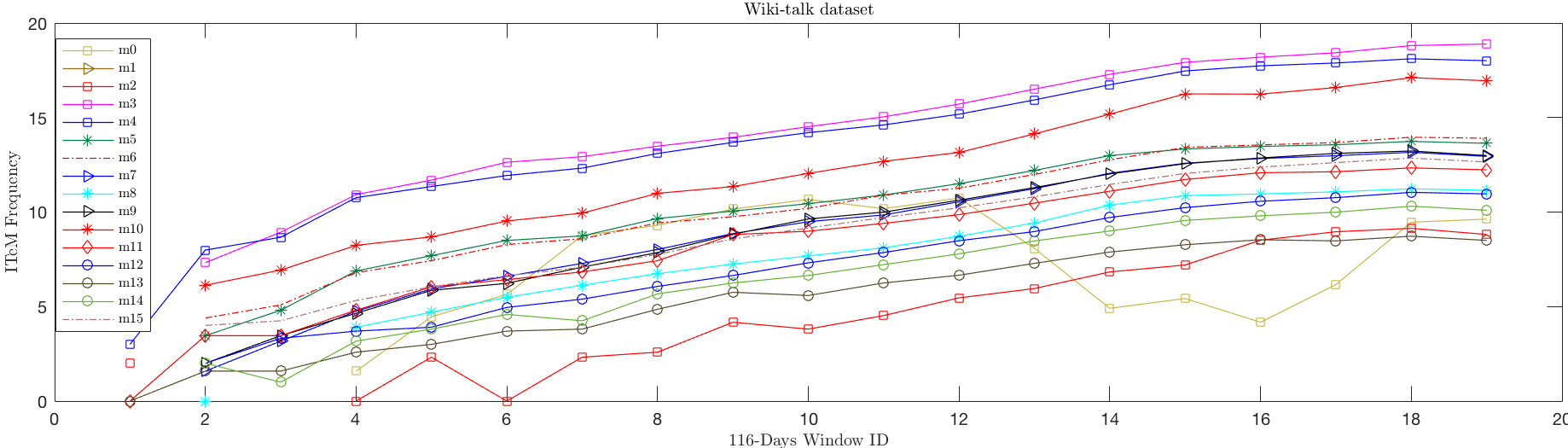}
	\vspace{-.7cm}
	\caption{ITeM frequency changes in the Wiki-talk (WT) temporal network }
	\label{fig:wiki_talk}
\end{figure*}

Figure \ref{fig:higgsresultInd} shows motif independence over time for the same window of the \textbf{HT}. Figures \ref{fig:higgsresult} and \ref{fig:higgsresultInd} show that the \textit{core} motif, such as the star, increases in count but the \textit{motif independence} decreases sharply. This happens as the temporal network exhibits the emergence of a hub-like structure with a small number of extremely-high degree vertices. In contrast to the \textit{burst} observed in the HT, Wiki-talk (\textbf{WT}) shows a linear evolution of the graph for a very long time (76 months) as shown in Figure \ref{fig:wiki_talk}.
\subsection{Scalability Analysis} 
A major contribution of this paper is a distributed algorithm to analyze a large temporal graph or a sequence of temporal graph windows. All the algorithms are developed using the Apache Spark 2.3.0, GraphFrame 0.7.0, and Scala 2.11.8 environment. This allows the use of scalable distributed data structures to handle large graphs in the order of millions of edges and to iteratively update the temporal-structural and orbital properties of the graph. To analyze the scalability of the core algorithm, we use a Snakemake \cite{koster2012snakemake} based automation pipeline and a SLURM \cite{yoo2003slurm} based resource manager. We experiment with different combinations of hardware resources and distributed partitions. Figure \ref{fig:emailScalability} shows the results of the scalability experiment using the EmailEU dataset. ITeM shows initial speed-up up to a maximum of 32 cores available to the Spark application. Beyond this point, the application suffers from communication and data serialization overhead. A similar trend was observed as we increased the number of data partitions, keeping the maximum number of cores fixed. The run-time sharply decreases as we increase the executor memory from 2GB to 6GB, and the decrease slows down after that.

Temporal analysis of an evolving network using a window-based approach poses memory constraints and scalability challenges as the number of windows increases. We preserve minimum information across the windows to maintain a global summary of the temporal network and to save window-specific summaries and vertex features to files, to be used by other analytic processes. This allows us to use our method in a longer running streaming fashion. Although we do not observe a strong sub-linear trend as the windows progress, as shown in Figure \ref{fig:redditruntime}, further analysis of the window graph structure using ITeM suggests that the run times depend on both the window size and the \textit{fringe} structure of the graph. The runtime of Window 5 and 10 decreases even as the graph size increases because those windows have a higher number of multi-edges in comparison to windows of similar size, which leads to aggressive subgraph reduction while discovering larger motifs. Future work will perform a more detailed analysis of the impact of a specific ITeM count on the runtime. For all three approaches, overall run-time complexity depends on enumerating larger motifs in the network but $\delta$-Motif has developed a set of specialized algorithms that count certain motif classes faster. Similarly, DG uses \textit{constrained dynamic graphlet counting}, a modified counting process to examine fewer instances of a given dynamic graphlet. In contrast, ITeM uses a general purpose framework to discover temporal motifs. This leads to faster run-times for $\delta$-Motif and DG but ITeM provides a fault-tolerant framework to analyze large graphs. Future work will also develop specialized distributed algorithms to find certain classes of motifs instances.
\vspace{-.4cm}
\section{Conclusion and Future Work} \label{sec:conc}
Complex temporal networks are observed in the real world, and a better understanding of them is required to effectively handle real-world applications. We present Independent Temporal Motif (ITeM) as a building block to characterize temporal graphs. ITeM reveals many salient features of the temporal graph, such as its core structure, \textit{fringe} vertices and edges, temporal evolution, and uniqueness. Graphs from different domains are found to exhibit varied structural and temporal distributions. Likewise, graphs from similar domains are found to exhibit similar structural properties, but many of them show varied temporal characteristics. We use these observations to characterize individual graphs and define a metric to quantitatively measure the similarity among them. We also present the  \textit{importance sampling} based approach to analyze a large graph as a sequence of smaller windows. We use this to show a change in the distribution that exhibits a behavioral shift in the way entities interact in a transactional graph, such as a social network. 

The rate at which temporal motifs are formed can also be used to generate synthetic graphs that exhibit similar evolution as a given real-world graph, as shown in \cite{purohittemporal}. Additionally, these features can also be used in a diverse set of applications, such as approximate sub-graph matching, graph mining, and network embedding learning. We will compare ITeM to other temporal network embeddings to measure the benefits of ITeM over other approaches for use in such applications. Future work will also address scalability challenges by estimating the number of ITeMs using specialized algorithms for different motif classes and perform a sensitivity analysis of the sampling approach. 

%

\bibliography{sample-base}
\bibliographystyle{aaai}
\end{document}